\documentclass[journal,comsoc]{IEEEtran}
\usepackage{amsmath,amsfonts}
\usepackage{algorithmic}
\usepackage{algorithm}
\usepackage{array}
\usepackage[caption=false,font=normalsize,labelfont=sf,textfont=sf]{subfig}
\usepackage{textcomp}
\usepackage{stfloats}
\usepackage{url}
\usepackage{verbatim}
\usepackage{graphicx}
\usepackage{cite}
\usepackage{color}
\usepackage{multirow}
\usepackage{booktabs}

\newcommand{\tabincell}[2]{\begin{tabular}{@{}#1@{}}#2\end{tabular}}
\newcommand{\tnew}[1]{\textcolor{black}{#1}}
\newcommand{\tdel}[1]{\textcolor[RGB]{192,192,192}{#1}}

\begin{document}

\title{Check-Belief Propagation Decoding of LDPC Codes}

\author{Wu Guan,~\IEEEmembership{Member,~IEEE,}
	    and Liping Liang%,~\IEEEmembership{Member,~IEEE,}
        % <-this % stops a space
\thanks{Manuscript received 2022. This work was supported by the National Natural Science Foundation of China (Grant No. 62271069).}
\thanks{W. Guan and L. Liang are both with the School
	of Integrated Circuits, Beijing University of Posts and Telecommunications,
	Beijing 100876, China(e-mail: guanwu@bupt.edu.cn and liangliping@bupt.edu.cn).}% <-this % stops a space
}

%\markboth{IEEE Transactions on Communications}%,~Vol.~14, No.~8, August~2022}%
%{Shell \MakeLowercase{\textit{et al.}}: CBP Decoding of LDPC Codes}

\IEEEpubid{IEEE transactions on communications, DOI:10.1109/TCOMM.2023.3308155}

\maketitle

\begin{abstract}
Variant belief propagation (BP) algorithms are applied to low-density parity-check (LDPC) codes. 
However, conventional decoders suffer from a large resource consumption due to gathering messages from all the neighbour variable-nodes and/or check-nodes through cumulative calculations. In this paper, a check-belief propagation (CBP) decoding algorithm is proposed. Check-belief is used as the probability that the corresponding parity-check is satisfied. All check-beliefs are iteratively enlarged in a sequential recursive order, and successful decoding will be achieved after the check-beliefs are all big enough. \tnew{Compared to previous algorithms employing a large number of cumulative calculations to gather all the neighbor messages, CBP decoding can renew each check-belief by propagating it from one check-node to another through only one variable-node, resulting in a low complexity decoding with no cumulative calculations.}
%, and each check-belief can be transferred from one check-node to another through only one variable-node. All check-beliefs are iteratively enlarged in a sequential recursive order, and successful decoding will be achieved after the check-beliefs are all big enough. Compared to previous algorithms employing a large number of cumulative calculations, CBP decoding can renew messages through only two other nodes with no cumulative calculations, resulting in a low complexity decoding.
%\tcolor{Each check-belief combining any one variable-to-check (V2C) message from its neighbor variable-node can generate a check-to-variable (C2V) message. Each variable-node combineing any one C2V message from the neighbor check-node can generate a V2C message. Each V2C message combining its older one can renew the check-belief. }
%\tcolor{Different from previous BP algorithms gathering messages from all its neighbor node, the check-beliefs are renewed through one variable-node and one check-node at an acceptable speed, which results in a low complexity decoding with little performance loss.} 
The simulation results and analyses show that the CBP algorithm provides little error-rate performance loss in contrast with the previous BP algorithms, but consumes much fewer calculations and memories than them. It earns a big benefit in terms of complexity.

\end{abstract}


\section{Introduction}
\IEEEPARstart{L}{ow}-density parity-check (LDPC) codes are well known for their ability to approach near Shannon capacity limits at relatively low complexities using iterative decoding \cite{ref1}. Since their rediscovery in the 1990s, LDPC codes have been extensively used in various applications, such as digital video broadcasts (DVBs) \cite{ref2}, IEEE 802.11ad (WiGig) \cite{ref4} and 5G New Radio (NR)\cite{ref5}. Various decoding algorithms have been suggested to improve the performance of LDPC codes.

The conventional flooding belief propagation (FBP) algorithm, proposed in 1996, was the first successful soft decoding method \cite{ref6}. It performs message passing with variable-to-check (V2C) and check-to-variable (C2V) phases iteratively. In each phase, the messages are sent from all the variable-nodes (check-nodes) to the corresponding check-nodes (variable-nodes). 
\iffalse
Iterative decoding works by exchanging messages in a phase-by-phase manner. 
\fi
The FBP algorithm can fully propagate messages between all the nodes in the code graph and has an excellent error-correcting performance within an acceptable number of iterations. To reduce the complexity,
\iffalse
of the full connection of the code graph in the FBP algorithm, the flooding message passing is scheduled in a sequential order \cite{ref7} or in a semiparallel order with quasi-cyclic code graphs \cite{ref8}. This balances throughput and complexity. Meanwhile, there are
\fi
various simplified FBP algorithms are presented, such as the min-sum \cite{ref9}, offset min-sum or normalized min-sum algorithms\cite{ref10}. 
\iffalse
These methods decrease the decoding complexity. 
\fi
They were widely used in many early coding systems.

Later, in 2004, the most popular decoding method for LDPC codes, named layered BP (LBP) algorithm, was presented \cite{ref12}. Different from the full parallel message exchange between phases in the FBP algorithm, messages are interactive between layers, and each layer corresponds to a check-node in LBP decoding. 
\iffalse
This method sequentially propagates the messages layer by layer, repeatedly reproducing the extrinsic information in the same iteration. It promotes the decoding convergence speed \cite{ref13}. 
Compared with the FBP algorithm, 
\fi
The LBP method can reduce the number of iterations by approximately half that of FBP while maintaining the same processing complexity \cite{ref13}. Many variants of LBP, including row message passing \cite{ref15}, column message passing \cite{ref16}, row-layered message passing \cite{ref17} and so on, were proposed during this period. The resulting convergence performance, combined with min-sum simplifying methods, makes LBP the main decoding algorithm in various LDPC code applications \cite{ref18}.

\IEEEpubidadjcol

To further reduce the number of iterations, a reliability-based scheduling method for BP decoding, named residual BP (RBP), was proposed in 2007 \cite{ref19}. In RBP, message passing is fully sequential, and messages are exchanged between edges in descending order of extrinsic information value (reliability). 
\iffalse
As the reliability of each edge changes in iterative decoding, the RBP algorithm should dynamically schedule the edge order to improve the extrinsic message updating speed. Thus, these methods are also referred to as informed dynamic scheduling (IDS) strategies \cite{ref20}. 
Usually, the C2V residual is used as the reliability measure, and an edge message with the maximum C2V residual is scheduled first. 
\fi
The error rate for RBP converges much faster than that of FBP and LBP.
\iffalse 
, but sometimes its performance is slightly worse than that of FBP and LBP because some edges may never be updated in the iterations. 
To address performance loss, silent-variable-node-free RBP (SVNF-RBP), which can update the messages of each variable-node equally, was proposed \cite{ref20}.
\fi 
Other methods, including silent-variable-node-free RBP (SVNF-RBP) \cite{ref20} and conditional innovation based RBP (CI-RBP)\cite{ref22}, were also presented to improve convergence. These reliability-based decoding strategies have attractive convergence speeds and error-rate performance, and they have become research focus areas.

\begin{figure}[!t]
	\centering
	\includegraphics[width=2.5in]{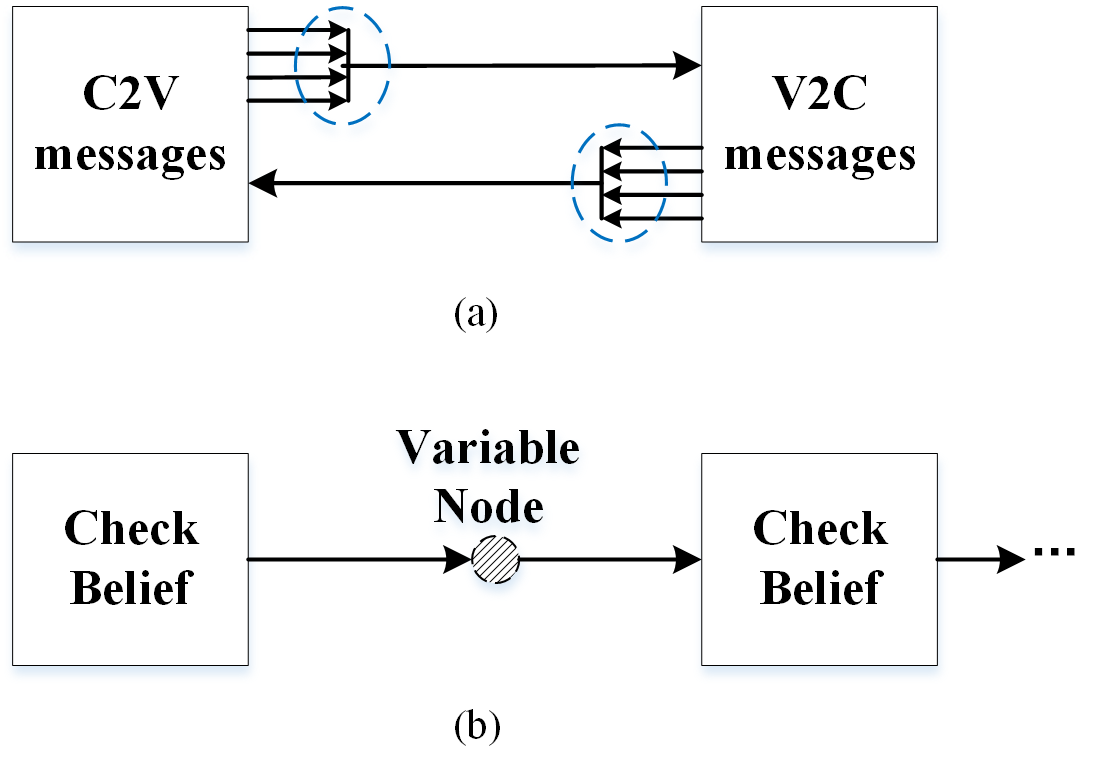}
	\caption{Message update methods of the previous algorithms and the CBP algorithm. (a) The previous algorithms gather messages from all the neighbor nodes through cumulative calculations. (b) The CBP algorithm renews check-belief from one check-node to another through only one variable-node with no cumulative calculations.}
	\label{fig_0}
\end{figure}

To decrease the complexity of LDPC decoding, the algorithms need to not only promote the convergence speed but also reduce the average calculation complexity of each message update. \tnew{As discussed above, the convergence speeds of the FBP, LBP and RBP increase in order. However, in these algorithms, as shown in Fig. \ref{fig_0}(a), the V2C and C2V messages are generated from the sum/product of all the other neighbour edges' messages of the variable-node/check-node [2]. In LDPC codes, the neighbour edges for each node are usually much larger than one. This results in a large number of complicated cumulative calculations and memory consumption for message updating. }
\iffalse
The cumulative calculations in edge message updates usually make the decoding complexity increase rapidly, especially for high-throughput applications. This will lead to large bottlenecks for decoding implementations. 
\fi
It usually makes the decoding complexity increase rapidly, and lead to large bottlenecks for  implementations.

To further reduce the complexity, LDPC decoding with no cumulative calculations has been studied.
\iffalse
To further reduce the complexity of the message calculations, iterative decoding is proposed to enlarge check-beliefs, satisfying the corresponding code parity checks. 
\fi
\tnew{As shown in Fig. \ref{fig_0}(b), each check-belief propagates between two check-nodes through only on variable-node. The CBP decoding updates the check-beliefs serially. Compared to other algorithms gathering messages from all the neighbor nodes, CBP decoding renews each check-belief through only two other nodes. This makes that there are no cumulative calculations in CBP. It results in reducing calculations and memories for in-row/in-column message scheduling, and earns a low complexity decoding.} Simulation results show that the CBP algorithm has little performance loss in contrast with the previous introduced algorithms. The analyses show that CBP consumes only one in hundreds of the calculation resources of RBP, SVNF-RBP and CI-RBP, much less calculation resources than FBP and LBP. Meanwhile, compared with the previous algorithms, it can reduce a large number of registers. Altogether, it has a much lower complexity than the above mentioned algorithms, making it suitable for high-throughput applications.

The rest of this paper is organized as follows. Section II reviews the FBP, LBP and RBP scheduling algorithms. Section III analyses the influence of cumulative calculations in variant BP decoding. Section IV proposes the CBP decoding algorithm. The error-rate performance and complexity of the CBP algorithm are investigated in Section V. Finally, Section VI concludes this paper.

\section{Preliminaries}
A binary $(N,K)$ LDPC code $C$ with rate $R=K/N$ is characterized by a code graph $\mathbf{G} = (\mathbf{V}, \mathbf{C}, \mathbf{E})$, where $\mathbf{V}$, $\mathbf{C}$ and $\mathbf{E}$ are the set of variable-nodes, check-nodes and edges, respectively. There are $N$ variable-nodes in $\mathbf{V}$ and $M=N-K$ check-nodes in $\mathbf{C}$. Let $N(v)$ and $N(c)$ denote the neighbour check-nodes of variable-node $v$ and  neighbour variable-nodes of check-node $c$, respectively. The symbols $N(v)\backslash c$ and $N(c)\backslash v$ denote the set $N(v)$ except for check-node $c$ and the set $N(c)$ except for variable-node $v$, respectively. 
\iffalse
The degrees of the variable-nodes with $i$ neighbours and the check-nodes with $j$ neighbours are denoted as $d_v^i$ and $d_c^j$, respectively. 
\fi
Its degree distributions are represented by $(\lambda(x) = \sum_i \lambda_i x^{i-1}, \rho(x) = \sum_j \rho_j x^{j-1})$, where $\lambda_i$ and $\rho_j$ denote the percentages of edges with degrees $d_v^i$ and $d_c^j$, respectively. The total number of edges in $\mathbf{E}$ is $E = N/(\sum_i \lambda_i/i) = M/(\sum_j \rho_j/j)$. 
\iffalse
The code graph can also be presented by an $M \times N$ parity-check matrix $\mathbf{H}=[h_{mn}]$, where each entry $h_{mn}=1$ corresponds to an edge connected by variable-node $v_n$ and check-node $c_m$. 
\fi
Let $\mathbf{X}=[x_v]$ and $\mathbf{Y}=[y_v]$ be the code transmitted and signal received, respectively.

\subsection{FBP decoding}

The FBP algorithm provides full parallel decoding for LDPC codes \cite{ref2}. 
\iffalse
It updates the V2C and C2V messages in two distinct phases and exchanges messages between the phases. 
The messages are propagated in a full parallel manner. 
The detailed message passing process is as follows.

• Step 1: Initialization.

The prior information $L_{v_a}$ for each variable-node $v_a \in \mathbf{V}$ and the C2V message $R_{c_i \rightarrow v_a}$ for each edge connected from check-node $c_i$ to variable-node $v_a$ are initialized as follows:
\begin{equation}
	\label{equ_s2e1}
	L_{v_a} = \ln \frac{Pr(y_{v_a}|x_{v_a}=0)}{Pr(y_{v_a}|x_{v_a}=1)}
\end{equation}
\begin{equation}
	\label{equ_s2e2}
	R_{c_i \rightarrow v_a} = 0
\end{equation}

• Step 2: V2C message update.
\fi
In the decoding, the V2C message sent from variable-node $v_a$ to check-node $c_i$ is updated as follows, 
\begin{equation}
	\label{equ_s2e3}
	Q_{v_a \rightarrow c_i} = L_{v_a} + \sum_{c_j \in N(v_a) \backslash c_i} R_{c_j \rightarrow v_a} 	
\end{equation}

\iffalse
• Step 3: C2V message update.
\fi

The C2V message sent from check-node $c_i$ to variable-node $v_a$ is calculated by
\begin{equation}
	\label{equ_s2e4}
	R_{c_i \rightarrow v_a} \!=\! \prod_{v_b \in N(c_i) \backslash v_a} \! \text{sgn} \left(Q_{v_b \rightarrow c_i} \right) \! \phi \left( \sum_{v_b \in N(c_i) \backslash v_a} \phi \left(\left| Q_{v_b \rightarrow c_i} \right| \right) \right)
\end{equation}
where
\begin{equation}
	\label{equ_s2e5}
	\phi(x)= -\log \left( \tanh \frac{x}{2} \right)
\end{equation} 

\iffalse
• Step 4: Posterior information generation.
\fi

The posterior information for each variable-node $v_a \in \mathbf{V}$
\begin{equation}
	\label{equ_s2e6}
	\Lambda_{v_a} = L_{v_a} + \sum_{c_j \in N(v_a)} R_{c_j \rightarrow v_a} 	
\end{equation}
is used to make tentative decision $\hat{\mathbf{X}}=[\hat{x}_{v_a}]$, where
\begin{equation}
	\label{equ_s2e7}
	\hat{x}_{v_a} = \begin{cases}
		0, & \Lambda_{v_a} \geq 0\\
		1, & \Lambda_{v_a} < 0\\
	\end{cases}	
\end{equation}

\iffalse
If $\mathbf{H} \hat{\mathbf{X}}^{T}=0$, or the maximum number of iterations is reached, stop decoding. Otherwise, iterative decoding goes on.

Notably, the stop criterion does not imply that there are no errors after decoding \cite{ref26}. Other frame error checking methods such as cyclic redundancy check (CRC), can also be used \cite{ref27}. However, most frame errors can be checked by the above stop criterion in low complexity. Thus, this basic stop criterion is used for the iteration.
\fi

From the above algorithm, we can see that FBP schedules all the V2C updates using (\ref{equ_s2e3}) in the first phase and all the C2V updates using (\ref{equ_s2e4}) in the second phase. Therefore, in each iteration, each edge transfers its extrinsic message to its neighbours in each update. This results in a full parallel but slow-message-transfer strategy.

\subsection{LBP decoding}
Different from the full parallel updating in FBP, 
\iffalse
LBP uses a sequential schedule to improve the message transfer depth. 
\fi
LBP decoding splits each iteration into multiple propagation layers. 
\iffalse
Each layer sequentially transfers its extrinsic messages to the neighbouring layers. 
\fi
This results in deep layer-by-layer message propagation in each iteration and hence promotes convergence.

In each layer, the V2C message is derived from the posterior information as
\begin{equation}
	\label{equ_s2e10}
	Q_{v_a \rightarrow c_i} = \Lambda_{v_a} - R_{c_i \rightarrow v_a}
\end{equation}
and the posterior information generated from the renewed C2V messages is
\begin{equation}
	\label{equ_s2e12}
	\Lambda^{new}_{v_a} = Q_{v_a \rightarrow c_i}  + R^{new}_{c_i \rightarrow v_a}
\end{equation}

\iffalse
In FBP, each new message is generated after all the neighbouring messages have been updated. 
\fi
In LBP decoders, each node updates its message using all its renewed neighbours and propagates its newest message to others in each iteration. This promotes the depth of message exchange in each iteration and hence speeds up convergence.

\subsection{RBP decoding}
\iffalse
In FBP and LBP, the messages are updated in a predefined order. 
\fi
To improve  convergence, RBP dynamically schedules the message with the maximum C2V residuals to be updated \cite{ref19}. This results in an edge-by-edge update process. 

The C2V residual is generated by the magnitude difference between the current C2V message $R_{c_i \rightarrow v_a}$  and the precomputed message $R_{c_i \rightarrow v_a}^{pre}$. The C2V messages $R_{c_i \rightarrow v_a}^{pre}$ are precomputed by (\ref{equ_s2e4}) for all the edges, and the corresponding C2V message residuals are generated by
\begin{equation}
	\label{equ_s2e13}
	r_{c_i \rightarrow v_a} = | R_{c_i \rightarrow v_a}^{pre} - R_{c_i \rightarrow v_a} |
\end{equation}

The RBP schedules the edge with the maximum C2V residual to be updated, and each updated C2V message is propagated to its neighbours. In this way, the factor for check-node updates is promoted from all the neighbours’ renewal of LBP to every neighbour’s renewal in RBP, which results in a more frequent message exchange and significantly increases the convergence speed. Other methods, including NW-RBP, SVNF-RBP and CI-RBP, can also improve convergence based on reliability-based scheduling.

\section{Influence of Cumulative Calculations in Decoding}
\subsection{Sole Edge Scheduling}
Sole edge scheduling is usually used in informed dynamic strategies for reliability-based BP decoding, such as RBP, SVNF-RBP and CI-RBP. In this type of decoding, message updating is scheduled in an edge-by-edge manner. Each schedule updates the sole C2V message. 
\iffalse
There are no relationships between the two subsequent updating edges. 
\fi
Both V2C and C2V updates gather messages from all the neighbours; thus, their calculations are based on the cumulative sum or cumulative product of their neighbours’ messages, as described in (\ref{equ_s2e3}) and (\ref{equ_s2e4}). Therefore, the calculation complexity of each update is $d_v^i-1$ for the V2C message updating of a variable-node with degree $d_v^i$, and $d_c^j-1$ for the  C2V message updating of a check-node with degree $d_c^j$. As $d_v^i-1$ and $d_c^j-1$ are no less than one, 
\iffalse
the complexity of decoding reflects a large number of message updates. Thus, the cumulative calculations
\fi
it results in a large complexity increase.
\iffalse
in the sole edge scheduling. This causes a large implementation bottleneck.
\fi
Therefore, the sole edge scheduling, such as RBP, SVNF-RBP and CI-RBP decoding, is not a popular method for high-throughput applications.

\subsection{Row/Column Scheduling}
Row scheduling or column scheduling, such as FBP and LBP, are popular methods for low-complexity BP decoding. In these methods, the message updating in (\ref{equ_s2e3}) and (\ref{equ_s2e4}) are updated as follows:
\begin{equation}
	\label{equ_s3e1}
	Q_{v_a \rightarrow c_i} = \left[L_{v_a} + \sum_{c_j \in N(v_a)} R_{c_j \rightarrow v_a} \right] - R_{c_i \rightarrow v_a} 	
\end{equation}
\begin{equation}
	\label{equ_s3e2}
	\begin{split}
		R_{c_i \rightarrow v_a} = & \left[\prod_{v_b \in N(c_i)} \text{sgn} \left(Q_{v_b \rightarrow c_i} \right) \right] \cdot \text{sgn} \left(Q_{v_a \rightarrow c_i} \right) \\
		& \cdot \phi \left( \left[\sum_{v_b \in N(c_i)} \phi \left(\left| Q_{v_b \rightarrow c_i} \right| \right) \right] - \phi \left(\left| Q_{v_a \rightarrow c_i} \right| \right) \right)
	\end{split}
\end{equation}

In this updating process, the values in $[\cdot]$ are shared. This means that for each variable-node $v_a \in N(c_i)$, the message updating of $Q_{v_a \rightarrow c_i}$ for all $c_i \in N(v_a)$ uses a shared sum $[L_{v_a} + \sum_{c_j \in N(v_a)} R_{c_j \rightarrow v_a} ]$. Similarly, a shared product is applied in the check-node processing, as shown in (\ref{equ_s3e2}). The output value is obtained from the cumulative result exclusive of the original message.

\begin{figure}[!t]
	\centering
	\includegraphics[width=3.1in]{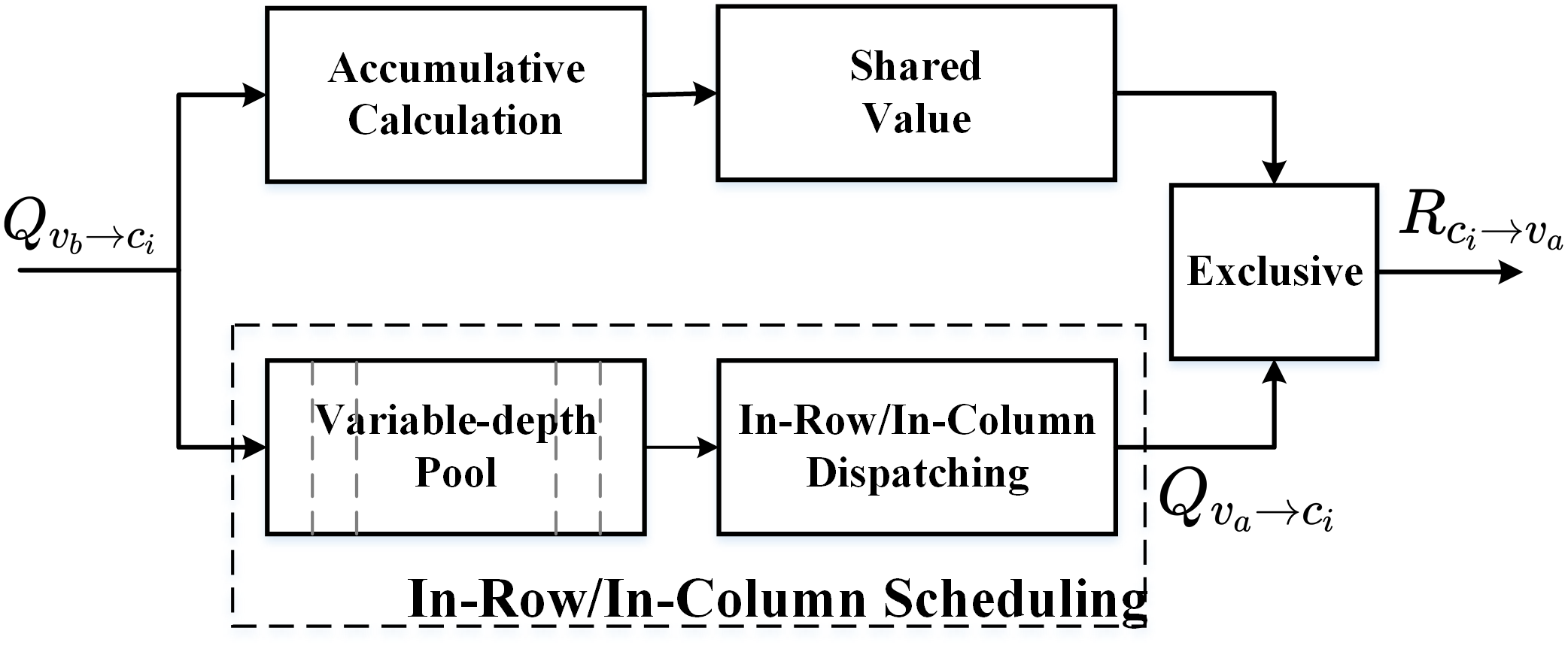}
	\caption{In-row/in-column scheduling for cumulative calculations.}
	\label{fig_1}
\end{figure}

In this updating, besides the conventional accumulative calculation for the shared value, an additional in-row/in-column scheduling process is needed to provide the original messages for the exclusive operation, as shown in Fig. \ref{fig_1}. The shared cumulative value in $[\cdot]$ can be generated recursively, and the in-row/in-column scheduling process is used to reserve the original messages and dispatch them serially when the shared value is generated. This structure can renew one extrinsic message in each unit time. It can obtain the highest throughput among the previous various decoding methods \cite{ref11}.

However, the in-row/in-column message scheduling consumes a large number of resources. First, a pool is needed for reserving the $d_v^i$ or $d_c^j$ messages. As the degrees of reserving messages varies in a wide range, the pool depth should be a variable. It should be pointed out that a variable-depth memory cell (i.e. register) takes up to 10 to 20 times area of a general memory cell to realize its variable-connections \cite{ref28}. The pool would consume many resources. Secondly, dispatching process is needed to select the original messages for the exclusive operation. Thirdly, a large number of in-row/in-column scheduling processes should be needed for parallel message updates in high-throughput applications. \\

%This is because the corresponding data lengths in each cumulative calculation vary over a large range. 
\iffalse
For example, for parallel decoding of LDPC codes in 5G NR, 384 parallel cumulative calculation units are needed for approximately 2 Gbps throughput with a maximum of 8 iterations \cite{ref23}. Meanwhile, the number of elements in each cumulative calculation, which is determined by the degree $d_c^j$, ranges from 3 to 19 \cite{ref5}. Thus, the pool depth is 19. Additionally, each message is soft quantized by 8 bits. Here, there are $384 \times 19 \times 8 = 58,368$ registers for the variable-depth pool. However, there are only $384 \times 8 = 3,072$ registers for C2V/V2C calculations. Thus, the additional registers used for cumulative calculations are more than nineteen times those used for message updating. This results in a large resource consumption. 
\fi

As discussed above, the cumulative calculations are highly complex, either from a large number of distributed calculations in the sole edge scheduling or from a large amount of resource consumption for the in-row/in-column scheduling. Moreover, a large number of cumulative calculations consume a large number of registers. It would take 10 to 20 times area of general memories. This is a very critical drawback for high-throughput systems. The aim of non-cumulative calculation algorithms is to decrease the complexity of LDPC decoding.

\section{check-belief Propagation Decoding}
\subsection{Check-beliefs Indicate the Probability of Successful Decoding}
In conventional decoding, the V2C and C2V messages are generated with the cumulative calculations of the neighbouring nodes’ messages. This causes the problem of complexity enlargement.

\begin{figure}[!t]
	\centering
	\includegraphics[width=3.1in]{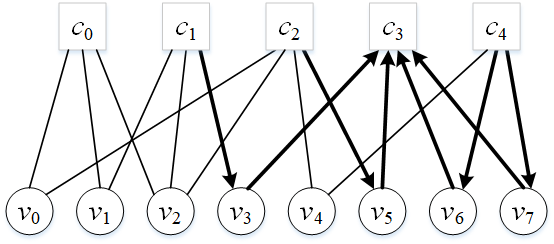}
	\caption{check-beliefs transferred from other check-nodes.}
	\label{fig_2}
\end{figure}

To avoid this problem, we decode the code with message exchanges between the check-nodes.
\iffalse
through a variable-node. 
\fi
Here, the check-belief of the check-node is denoted as follows \cite{ref29}:
\begin{equation}
	\label{equ_s4e1}
	\Omega_{c_i} = \log \left( \frac{Pr(S_{c_i} = 0|\mathbf{Y})}{Pr(S_{c_i} = 1|\mathbf{Y})} \right)
\end{equation}
where $S_{c_i}$ denotes the parity check corresponding to check-node $c_i$.

The check-belief denotes the probability that the parity check of the check-node is satisfied. For a satisfied parity check, the check-belief should be a positive value; otherwise, it should be a negative value. Thus, for successful decoding, check-beliefs for all check-nodes should be positive, that is,
\begin{equation}
	\label{equ_s4e2}
	\Omega_{c_i} > 0, \forall c_i \in \mathbf{C}
\end{equation}

The check-beliefs can be propagated. As shown in Fig.\ref{fig_2}, in the code graph, the check-beliefs transferred to check-node $c_3$ are from check-node $c_1$ through $v_3$, from check-node $c_2$ through $v_5$, from check-node $c_4$ through $v_6$ and from check-node $c_4$ through $v_7$. Therefore, the check-belief updating of check-node $c_3$ is based on the check-beliefs of check-nodes $c_1$, $c_2$ and $c_4$. Each check-node can receive several check-beliefs from the other check-nodes. This can help the check-node to increase its check-belief. By iteratively propagating the check-beliefs among the code graphs, all the check-beliefs for the check-nodes can be promoted to positive values, and thus successful decoding is achieved.

\subsection{Check-Node to Check-Node Check-Belief Propagation}
check-beliefs propagate from one check-node to other check-nodes through variable-nodes. They can be generated as follows \cite{ref2}:
\begin{equation}
	\label{equ_s4e3}
	\Omega_{c_i} = \prod_{v_a \in N(c_i)} \text{sgn} \left(Q_{v_a \rightarrow c_i} \right) \cdot \phi \left( \sum_{v_a \in N(c_i)} \phi \left(\left| Q_{v_a \rightarrow c_i} \right| \right) \right)
\end{equation}

According to the definition in (\ref{equ_s4e3}), for every new check-node, its check-belief $\Omega_{c_i}^{new}$ can be updated by renewing each of its V2C messages $Q^{new}_{v_a \rightarrow c_i}$ from its neighbouring variable-nodes as follows,
\begin{equation}
	\label{equ_s4e4}
	\Omega_{c_i}^{new} = \prod_{v_a \in N(c_i)} \text{sgn} \left(Q_{v_a \rightarrow c_i}^{new} \right) \cdot \phi \left( \sum_{v_a \in N(c_i)} \phi \left(\left| Q_{v_a \rightarrow c_i}^{new} \right| \right) \right)
\end{equation}

As proven in Appendix A, the check-belief in (\ref{equ_s4e4}) can also be calculated in a recursive manner, that is,
\begin{equation}
	\label{equ_s4e5}
	\Omega_{c_i}^{(n)} = \psi^{+} \left(\Omega_{c_i}^{(n-1)}, Q_{v_{a_n} \rightarrow c_i}^{new} 	\right)
\end{equation}
where $n=0,1,\cdots,|N(c_i)|-1$, $|N(c_i)|$ is the number of elements in set $N(c_i)$, and
\begin{equation}
	\label{equ_s4e6}
	\psi^{+} (x,y) = \text{sgn}(x) \cdot \text{sgn}(y) \cdot \phi \left( \phi (|x|) + \phi(|y|) \right)
\end{equation}

Here, $\Omega_{c_i}^{(-1)}$ is initialized as $\Omega_{c_i}^{(-1)} = \infty$ and
\begin{equation}
	\label{equ_s4e8}
	\Omega_{c_i}^{new} = \Omega_{c_i}^{(\left| N(c_i) \right| - 1)}
\end{equation}

As calculated in (\ref{equ_s2e10}), the extrinsic V2C message $Q^{new}_{v_a \rightarrow c_i}$ can be obtained from the posterior information exclusive to the prior C2V message through variable-node $v_a$ as follows:
\begin{equation}
	\label{equ_s4e9}
	Q^{new}_{v_a \rightarrow c_i} = \Lambda_{v_a}^{new} - R_{c_i \rightarrow v_a}
\end{equation}

\begin{figure}[!t]
	\centering
	\includegraphics[width=2.3in]{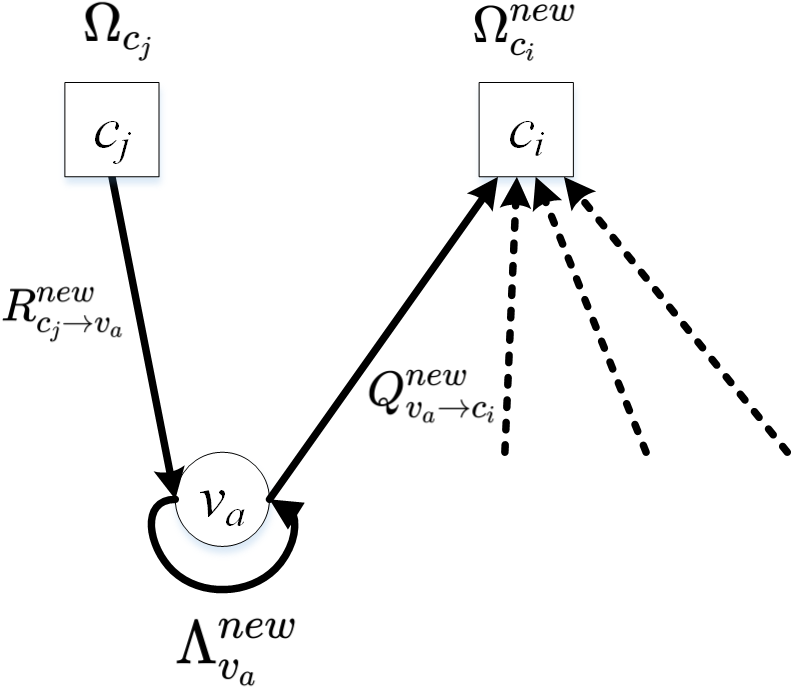}
	\caption{Check-node to check-node check-belief propagation.}
	\label{fig_3}
\end{figure}

Furthermore, by combining (\ref{equ_s2e10}) and (\ref{equ_s2e12}), we can see that the posterior information in (\ref{equ_s4e9}) is renewed by the update of another neighbouring check-node that is different from check-node $c_i$, denoted as $c_j$. This process is given as follows:
\begin{equation}
	\label{equ_s4e10}
	\Lambda_{v_a}^{new} = \Lambda_{v_a} - R_{c_j \rightarrow v_a} + R_{c_j \rightarrow v_a}^{new}
\end{equation}

The C2V message $R_{c_j \rightarrow v_a}^{new}$ in (\ref{equ_s4e10})
\iffalse
which is updated from check-node $c_j$ to variable-node $v_a$, 
\fi
can be obtained from the check-belief of $c_j$ by excluding its prior message following (\ref{equ_s3e2}), that is,
\begin{equation}
	\label{equ_s4e11}
	\begin{split}
		R^{new}_{c_j \rightarrow v_a} = & \left[\prod_{v_b \in N(c_j)} \text{sgn} \left(Q_{v_b \rightarrow c_j} \right) \right] \cdot \text{sgn} \left(Q_{v_a \rightarrow c_j} \right) \\
		& \cdot \phi \left( \left[\sum_{v_b \in N(c_j)} \phi \left(\left| Q_{v_b \rightarrow c_j} \right| \right) \right] - \phi \left(\left| Q_{v_a \rightarrow c_j} \right| \right) \right)
	\end{split}
\end{equation}

As proven in Appendix A, by combining (\ref{equ_s4e3}) and (\ref{equ_s4e11}), we can obtain
\begin{equation}
	\label{equ_s4e12}
	R_{c_j \rightarrow v_a}^{new} = \psi^{-} \left(\Omega_{c_j}, Q_{v_{a} \rightarrow c_j} 	\right)
\end{equation}
where
\begin{equation}
	\label{equ_s4e13}
	\psi^{-} (x,y) = \text{sgn}(x) \cdot \text{sgn}(y) \cdot \phi \left( \left| \phi (|x|) - \phi(|y|) \right| \right)
\end{equation}

Here, the prior V2C message $Q_{v_{a} \rightarrow c_j}$ is obtained from the original posterior information following (\ref{equ_s2e10}), that is,
\begin{equation}
	\label{equ_s4e14}
	Q_{v_a \rightarrow c_j} = \Lambda_{v_a} - R_{c_j \rightarrow v_a}
\end{equation}

The above check-belief updating process is shown in Fig.\ref{fig_3}. In this process, first, with the original check-belief $\Omega_{c_j}$ of check-node $c_j$, we can update the C2V message $R^{new}_{c_j \rightarrow v_a}$ by (\ref{equ_s4e11}). Second, the corresponding posterior information $\Lambda^{new}_{v_a}$ of variable-node $v_a$ is updated by (\ref{equ_s4e10}). Third, the variable-node $v_a$ sends a new V2C message $Q^{new}_{v_a \rightarrow c_i}$ to check-node $c_i$, as described by (\ref{equ_s4e9}). Finally, check-node $c_i$ updates its check-belief $\Omega_{c_i}^{new}$ in a recursive manner, following (\ref{equ_s4e5}). In this way, check-node $c_j$ transfers its check-belief $\Omega_{c_j}$ to check-node $c_i$ through variable-node $v_a$. This process provides a way to exchange the check-beliefs between the check-nodes, which can promote the reliability of the check-beliefs, ensuring successful decoding.

\subsection{Check-belief Propagation Decoding}

\begin{figure}[!t]
	\centering
	\includegraphics[width=3.2in]{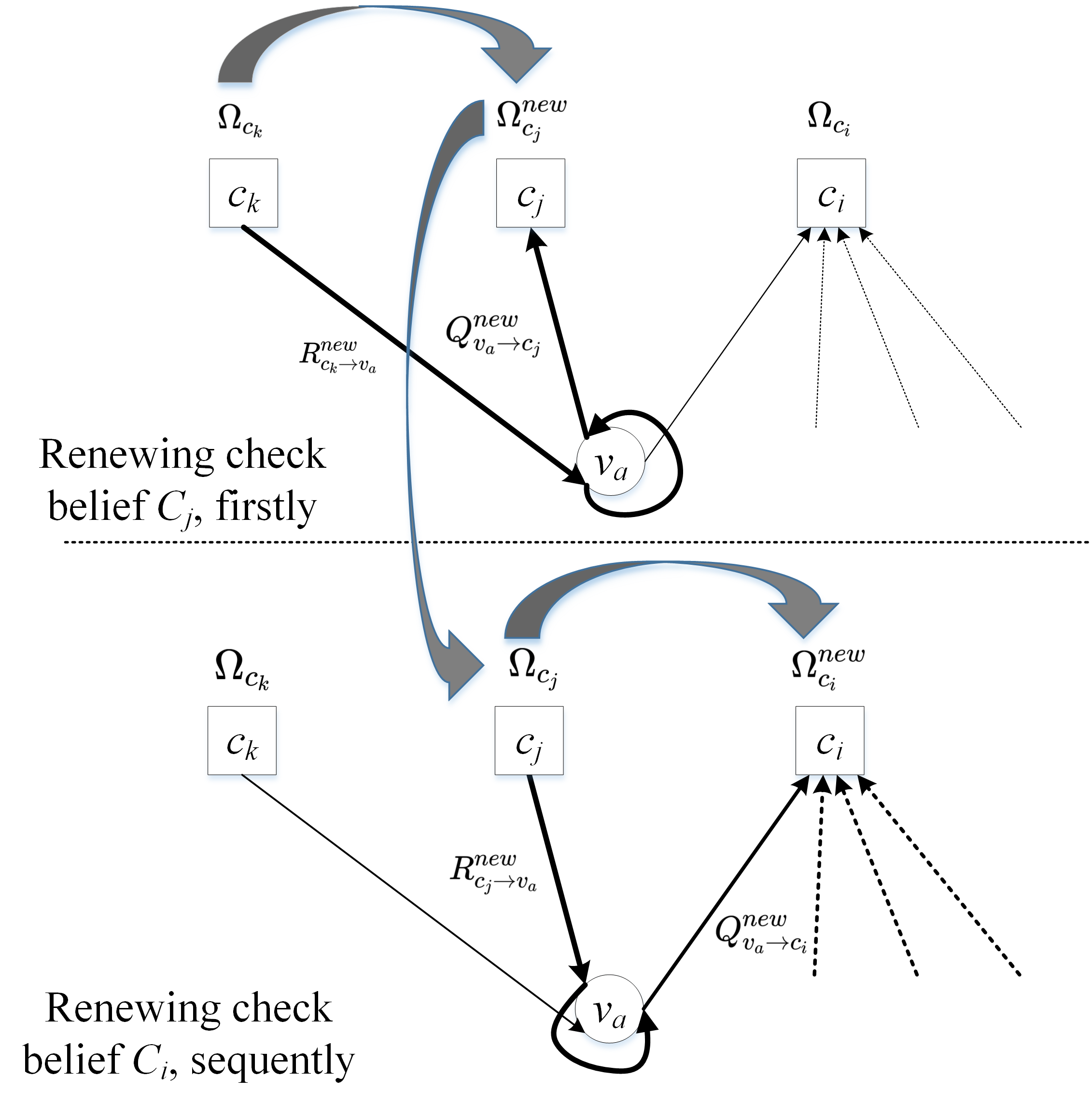}
	\caption{Sequentially check-belief propagation decoding.}
	\label{fig_4}
\end{figure}

Here, there are usually more than two neighbouring check-nodes connected to variable-node $v_a$. Thus, there are multiple check-beliefs transferred from the same variable-node, as shown in Fig.\ref{fig_4}. In this code, to update the check-belief $\Omega_{c_i}$ of check-node $c_i$ through $v_a$, there are two candidate check-nodes $c_j$ and $c_k$. Therefore, there are two candidate check-beliefs $\Omega_{c_j}$ and $\Omega_{c_k}$, which can both transfer their check-beliefs to check-node $c_i$ through variable-node $v_a$. For good message propagation, we transfer the check-beliefs in a sequential manner. As shown in Fig.\ref{fig_4}, firstly, the check-belief $\Omega_{c_k}$ is transferred to check-node $c_j$ through variable-node $v_a$, and new check-belief $\Omega^{new}_{c_j}$ is generated. Then, the new check-belief $\Omega^{new}_{c_j}$ is used as prior information $\Omega_{c_j}$, and transferred to check-node $c_i$ to generate the check-belief $\Omega^{new}_{c_i}$. In this way, the check-beliefs are sequentially updated.

From the above analysis, we can see that in sequential decoding, each check-belief $\Omega_{c_j}$ of check-node $c_j$ is used to update the check-belief of the nearest check-node $c_i$. Conversely, the check-belief of check-node $c_i$ is updated by the latest check-beliefs through its neighbouring variable-nodes. This means that for each neighbouring variable-node $v_a$ of check-node $c_i$, there are several neighbouring check-nodes $c_k,\cdots,c_j$, and thus they correspond to the  $N(v_a)$ V2C messages $R_{c_k \rightarrow v_a}, \cdots,R_{c_j \rightarrow v_a}$. For effective updating of the check-belief $\Omega_{c_i}$ through variable-node $v_a$, the latest check-belief $\Omega_{c_j}$ is used for sequential updating. The newest extrinsic information is propagated from this check-belief to other nodes, which results in improving the convergence speed for sequential decoding.

Meanwhile, only the latest V2C message is used in the sequential decoding. Thus, all the latest V2C messages can be denoted by the same symbol, $Q_{v_a \rightarrow c^{*}}$.

In this way, the CBP decoding algorithm can be described as follows.

• Step 1: Initialization.

For each variable-node $v_a \in \mathbf{V}$, the latest V2C message is
\begin{equation}
	\label{equ_s4e15}
	Q_{v_a \rightarrow c^{*}} = \ln \frac{Pr(y_{v_a}|x_{v_a}=0)}{Pr(y_{v_a}|x_{v_a}=1)}
\end{equation}

For each check-node $c_i$, the check-belief is
\begin{equation}
	\label{equ_s4e16}
	\Omega_{c_i} = \infty 
\end{equation}

For each edge connected from check-node $c_i$ to variable-node $v_a$, the C2V message is
\begin{equation}
	\label{equ_s4e17}
	R_{c_i \rightarrow v_a} = 0	
\end{equation}

Initialize the first check-node as the latest updated one.

• Step 2: Sequentially check the belief update.

For each check-node $c_i$, update its check-belief recursively.

(1) check-node to check-node check-belief propagation.

a) For each neighbouring variable-node $v_a \in N(c_i)$, the latest updated neighbouring check-node of variable-node $v_a$ is denoted as $c_j$.

b) Check-belief to variable-node (B2V) message updating following (\ref{equ_s4e12}),
\begin{equation}
	\label{equ_s4e18}
	R_{c_j \rightarrow v_a}^{new} = \psi^{-} \left(\Omega_{c_j}, Q_{v_a \rightarrow c^{*}} 	\right)
\end{equation}

c) V2C message updating. By combining (\ref{equ_s4e9}), (\ref{equ_s4e10}) and (\ref{equ_s4e14}), the V2C message is as follows:
\begin{equation}
	\label{equ_s4e19}
	Q_{v_a \rightarrow c^{*}}^{new} = Q_{v_a \rightarrow c^{*}} + R^{new}_{c_j \rightarrow v_a} - R_{c_i \rightarrow v_a}
\end{equation}

Meanwhile, according to (\ref{equ_s4e10}) and (\ref{equ_s4e14}), the posterior information update is generated as
\begin{equation}
	\label{equ_s4e20}
	\Lambda^{new}_{v_a} = Q_{v_a \rightarrow c^{*}} + R^{new}_{c_j \rightarrow v_a}
\end{equation}

It is one part of V2C message updating. Following (\ref{equ_s4e20}), the hard decision $\hat{x}_{v_a}$ can be obtained by (\ref{equ_s2e7}).

d) Following (\ref{equ_s4e5}), V2C message to check-belief (C2B) updating is as follows,
\begin{equation}
	\label{equ_s4e21}
	\Omega_{c_i}^{(n)} = \psi^{+} \left(\Omega_{c_i}^{(n-1)}, Q_{v_a \rightarrow c^{*}}^{new} 	\right)
\end{equation}
where $\Omega_{c_i}^{-1} = \infty$ and  $\Omega_{c_i}^{new} = \Omega_{c_i}^{|N(c_i)|-1}$.

e) Update the corresponding updated messages for check-node $c_i$ and variable-node $v_a$ as follows.
\begin{equation}
	\label{equ_s4e22}
	\begin{array}{ccc}
		R_{c_i \rightarrow v_a} & \leftarrow & R_{c_i \rightarrow v_a}^{new} \\
		Q_{v_a \rightarrow c^{*}} & \leftarrow & Q_{v_a \rightarrow c^{*}}^{new} \\
		%\Lambda_{v_a} & \leftarrow & \Lambda_{v_a}^{new} \\
		%\Omega_{c_i} & \leftarrow & \Omega_{c_i}^{new} \\
	\end{array}
\end{equation}

This means that the renewed messages in the updating of check-node $c_i$ are used as prior information for the updating of the next check-node. This results in sequential decoding.

(2) The updated check-belief is renewed as a prior belief as follows:
\begin{equation}
	\label{equ_s4e23}
	\Omega_{c_i}  \leftarrow  \Omega_{c_i}^{new}
\end{equation}

(3) Stopping criterion test.

If all the $N$ consecutive check-beliefs satisfy (\ref{equ_s4e2}) and the posterior information signs in (\ref{equ_s4e20}) have no flips in the $N$ consecutive check-belief updates, the decoding succeeds, that is,
\begin{equation}
	\label{equ_s4e24}
	\text{sgn}(\Lambda_{v_a}^{new}) = \text{sgn}(\Lambda_{v_a})
\end{equation}
Otherwise, go back to Step 2 until the maximum number of iterations is reached.

• Step 3: Output the hard decision $\hat{\mathbf{X}} = [\hat{x}_{v_a}]$ generated in the V2C phase of Step 2. \\

During decoding, each check-belief are transferred through B2V, V2C and C2B phases, following  (\ref{equ_s4e18}),  (\ref{equ_s4e19}) and  (\ref{equ_s4e21}), respectively.
\iffalse
combining one V2C message from its neighbouring variable-node can generate a C2V message following (\ref{equ_s4e18}). Each variable-node combining one C2V message from the neighboring check-node can generate a V2C message following (\ref{equ_s4e19}). Each V2C message combined its older message can update the check-belief following (\ref{equ_s4e21}). 
\fi
In this way, each check-belief can be transferred from one check-node to another check-node through only one variable-node. 
\iffalse
Thus, each check-belief can be updated by gathering messages from these nodes.
\fi
All the check-beliefs are iteratively enlarged in a sequential recursive order to ensure that all the parity checks are successfully satisfied. 
\iffalse
However, in conventional decoding, the messages are collected following (\ref{equ_s3e1}) and (\ref{equ_s3e2}). The messages from all the neighbouring variable and/or check-nodes are collected; thus, updating is achieved through	cumulative sums and products. There are cumulative calculations for each message updating, which makes decoding very complex.
\fi
Different from previous algorithms, each check-belief is propagated through two other nodes with no cumulative calculations, which results in low complexity decoding with little performance loss.

\iffalse
The detailed performance comparison and complexity analysis in Section V will confirm this result.
\fi
	
%In this way, each check-belief is updated in a recursive way, and all the check-beliefs are updated in a sequential order. Iterative check-belief propagation promotes all the check-beliefs and then results in a successful decoding. 
	
%Different from previous BP algorithms gathering messages from all its neighbor node, the check-beliefs are renewed through one variable-node and one check-node at an acceptable speed, which results in a low complexity decoding with little performance loss.} 

%In the decoding, each check-belief is updated in a recursive way, and all the check-beliefs are updated in a sequential order. Iterative check-belief propagation promotes all the check-beliefs and then results in a successful decoding. Different from the previous introduced algorithms, there are no accumulative calculations in this process, and the stopping criterion tests are generated directly from the check-beliefs. This decreases the decoding complexity.

\subsection{Straightforward Decoding Structure with No Cumulative Calculations}
\begin{figure}[!t]
	\centering
	\includegraphics[width=3.2in]{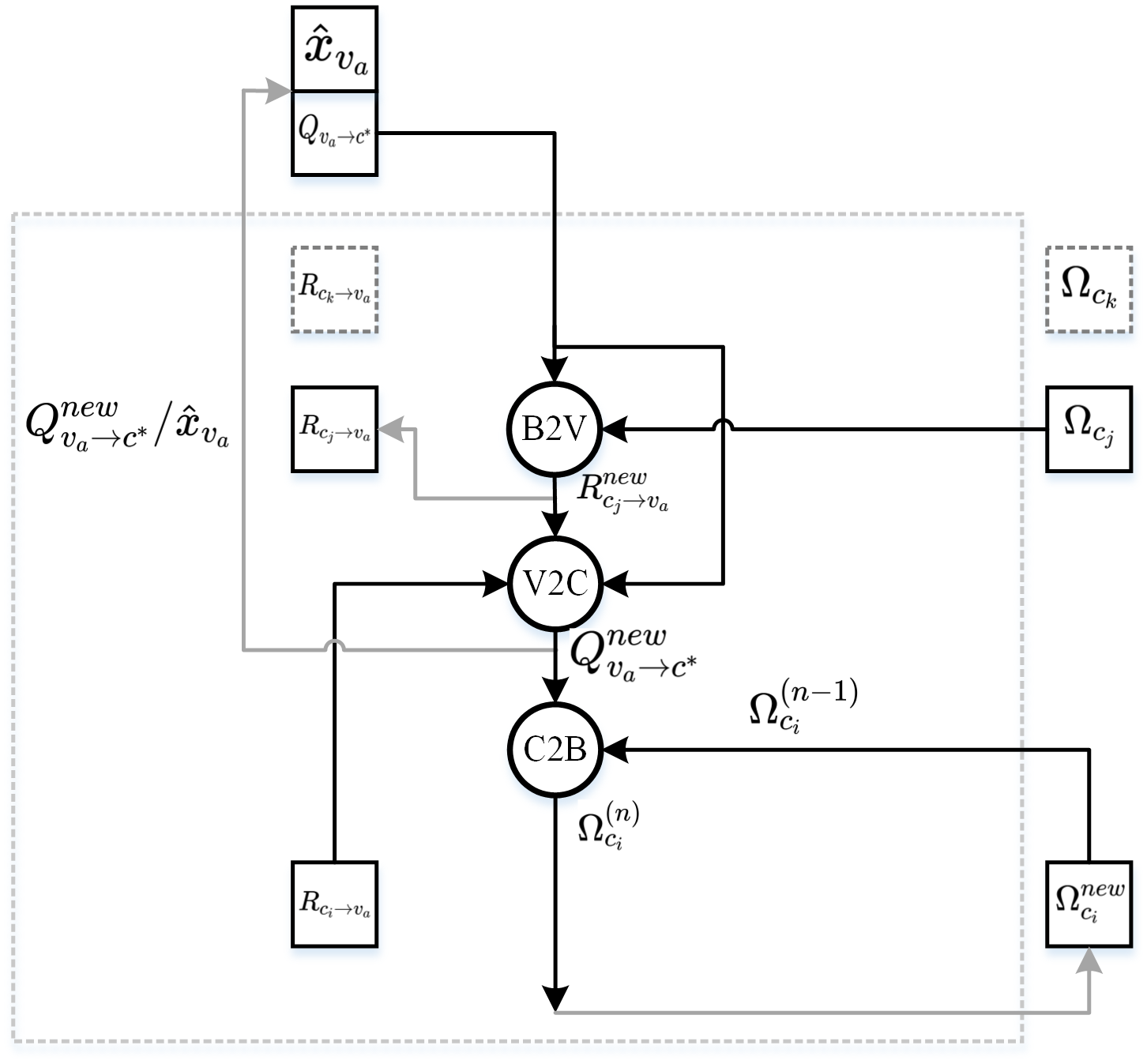}
	\caption{No cumulative calculation decoding structure for CBP decoding.}
	\label{fig_5}
\end{figure}

The straightforward decoding structure is shown in Fig.\ref{fig_5}. The check-beliefs $\Omega_{c_i}$  for all the check-nodes, the latest V2C messages $Q_{v_a \rightarrow c^{*}}$ for all variable-nodes and the C2V messages $R_{c_i \rightarrow v_a}$ for all the edges in the code graph are retained in each iteration.

In decoding, there are three phases: the B2V, V2C and C2B phases. Firstly, in the B2V phase, the check-belief $\Omega_{c_j}$ and the latest V2C message $Q_{v_a \rightarrow c^{*}}$ are collected, and the updated C2V message $R^{new}_{c_j \rightarrow v_a}$ is generated according to (\ref{equ_s4e18}). Secondly, in the V2C phase, the prior V2C message $Q_{v_a \rightarrow c^{*}}$, the prior C2V message $R_{c_i \rightarrow v_a}$ and the updated C2V message $R^{new}_{c_j \rightarrow v_a}$ are gathered, and the updating message $Q^{new}_{v_a \rightarrow c^{*}}$ is generated following (\ref{equ_s4e19}). Specifically, the decision of the variable-node, $\hat{x}_{v_a}$, can also be decided in this step. Thirdly, in the C2B phase, the check-belief is updated recursively using the renewed V2C messages $Q^{new}_{v_a \rightarrow c^{*}}$.

Here, we can see that there are no cumulative calculations in the three phases. 
\iffalse
There are also no conflicting updates in the three-phase decoding process. 
\fi
The three phases can be conducted in a straight pipeline.
\iffalse
without any additional resource consumption. 
\fi  
This results in a low-complexity decoding structure.
Meanwhile, 
\iffalse
in each pipeline, the messages will be transferred on the edge from the check-node to the variable-node and on the edge from the variable-node to the check-node; thus, 
\fi
messages are transferred by two edges in each check-belief renewing. This promotes the message propagation depth and results in a high convergence speed.

Furthermore, the B2V update in (\ref{equ_s4e18}) and C2B update in (\ref{equ_s4e21}) can be simplified based on a normalized min-sum approach:
\begin{equation}
	\label{equ_s4e18xx}
	R_{c_j \rightarrow v_a}^{new} = \left( \Omega_{c_j}^{min} == Q_{v_a \rightarrow c^{*}} 	\right) ? \Omega_{c_j}^{submin} : \Omega_{c_j}^{min}
\end{equation}
\iffalse
\begin{equation}
	\label{equ_s4e19xx}
	Q_{v_a \rightarrow c^{*}}^{new} = Q_{v_a \rightarrow c^{*}} + R^{new}_{c_j \rightarrow v_a} - R_{c_i \rightarrow v_a}
\end{equation}
\fi
\begin{equation}
	\label{equ_s4e21xx}
	\begin{array}{l}
		if \left(Q_{v_a \rightarrow c^{*}}^{new} < \Omega_{c_i}^{min}	\right)	\\
		\quad \quad \left(\Omega_{c_i}^{min}, \Omega_{c_i}^{submin} \right) = \left( \alpha \cdot Q_{v_a \rightarrow c^{*}}^{new}, \Omega_{c_i}^{min} \right)\\ 
		elseif \left(Q_{v_a \rightarrow c^{*}}^{new} < \Omega_{c_i}^{submin}	\right) \\
		\quad \quad \left(\Omega_{c_i}^{min}, \Omega_{c_i}^{submin} \right) = \left( \Omega_{c_i}^{min}, \alpha \cdot Q_{v_a \rightarrow c^{*}}^{new}  \right)\\ 
		end	
	\end{array} 
\end{equation}
where $\Omega_{c_i}^{min}$ and $\Omega_{c_i}^{submin}$ are the minimum and the sub-minimum value of the min-sum approximation, and $\alpha$ is the normalized parameter \cite{ref10}.

\iffalse
\tdel{Following the above process, CBP decoding can be simplified based on a normalized min-sum approach. }
\fi
In this way, CBP decoding can be implemented by \textit{min} and \textit{sum} functions instead of the \textit{log-tanh} functions, making it suitable for hardware implementation.

\section{Performance Simulation and Complexity Analysis}
We compare the performance of the traditional algorithms and the proposed CBP algorithm through the binary phase shift keying (BPSK) modulated  additive white Gaussian noise (AWGN) channel. Comparisons are made for the regular (3,6) LDPC codes and irregular LDPC codes under degree distributions ($\lambda(x)= 0.45 x^1 + 0.3708 x^2 + 0.0307 x^3 + 0.1485 x^{11}$, $\rho(x) = 0.5467 x^4 + 0.4533 x^5$). The parity-check matrices of the LDPC codes are constructed using the progress edge growth (PEG) algorithm \cite{ref24}. The codes simulated are all rate-1/2 LDPC codes. The maximum number of iterations is 200. 
\iffalse
It should be noted that the complexity is analyzed in low error-rate conditions, to match the excellent performance of LDPC codes.
\fi

\subsection{Error Correction Performance}
\iffalse
The bit error rate (BER)/frame error rate (FER) to signal-to-noise ratio (SNR) performances for the different decoding schemes are shown in Fig.\ref{fig_6} and Fig.\ref{fig_7}.
\fi

\begin{figure}[!t]
	\centering
	\subfloat[]{\includegraphics[width=1.7in]{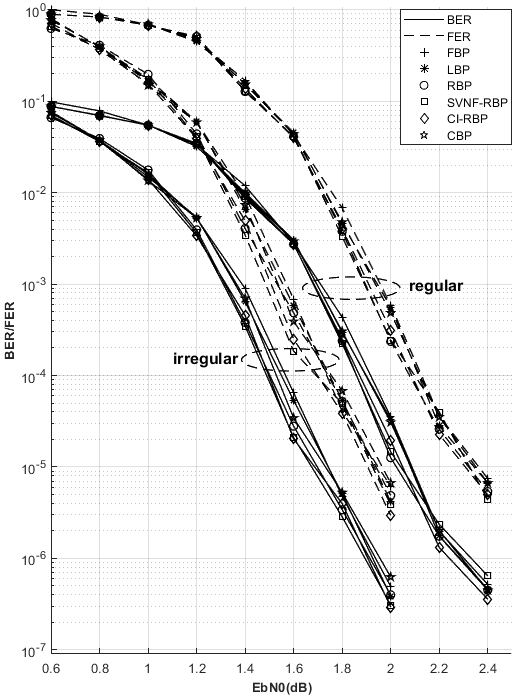}%
		\label{fig_6}}
	\hfil
	\subfloat[]{\includegraphics[width=1.7in]{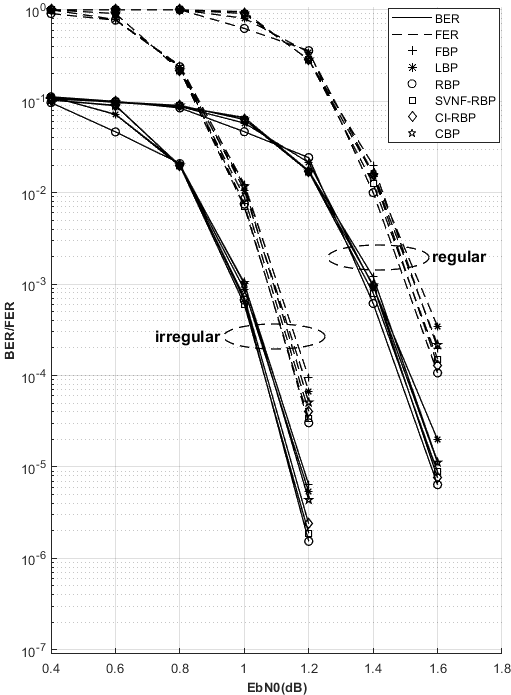}%
		\label{fig_7}}
	\caption{Performance comparisons of the FBP, LBP, RBP, SVNF-RBP, CI-RBP and the proposed CBP scheduling strategies. (a)  Length-2048 LDPC codes. (b) Length-8192 LDPC codes.}
	\label{fig_6_7}
\end{figure}

\iffalse
\begin{figure}[!t]
	\centering
	\includegraphics[width=3in,height=2.7in]{fig6_bfer_2048}
	\caption{Performance comparisons of the FBP, LBP, RBP, SVNF-RBP, CI-RBP and the proposed CBP scheduling strategies for length-2048 LDPC codes.}
	\label{fig_6}
\end{figure}

\begin{figure}[!t]
	\centering
	\includegraphics[width=3in,height=2.7in]{fig7_bfer_8192}
	\caption{Performance comparisons of the FBP, LBP, RBP, SVNF-RBP, CI-RBP and the proposed CBP scheduling strategies for length-8192 LDPC codes.}
	\label{fig_7}
\end{figure}
\fi

Fig.\ref{fig_6_7} show the AWGN performance of the FBP, LBP, RBP, SVNF-RBP, CI-RBP scheduling strategies discussed above and the proposed CBP strategy for regular and irregular LDPC codes, respectively. The figures show that the proposed CBP has a comparable performance to the other algorithms.
As indicated in Fig.\ref{fig_6_7}, the performance differences for FBP, LBP, RBP, SVNF-RBP, CI-RBP and CBP are less than 0.05 dB for the same code. This is because these decoding algorithms all use belief propagation decoding. 
\iffalse
The differences between them are the scheduling orders. While this can cause different convergence speeds, the total extrinsic information would be no changing. Thus, there should be no performance difference between these decoding strategies.
\fi
\tnew{They use different scheduling order with different message propagation depth. This can cause different convergence speeds. However, the extrinsic information is all generated from the same parity-checks. Thus, they have similar error-correcting performances. On the other hand, RBP, SVNF-RBP and CI-RBP can schedule the flipped nodes firstly. This would decrease the influence of trapping sets, and win a performance improvement \cite{ref20}. Especially, CI-RBP can conditionally schedule the flipped variable-nodes to control the influence of trapping sets, it owns much better performance than others\cite{ref22}.}

%Indeed, Fig.\ref{fig_6} shows that CBP has a better performance than FBP, and a slightly worse performance than RBP, SVNF-RBP and CI-RBP for high SNRs. This is because RBP decoding propagates the extrinsic messages deeply through the edges by edge scheduling, and the LBP and CBP decoding transfer their extrinsic messages layer to layer, with the sequential messages reused in iterations. Unfortunately, FBP decoding exchanges the extrinsic messages, phase by phase, which results in a slow message transferring effect. It can also be seen that CBP performs almost indistinguishably from LBP. This is because the proposed CBP exchanges the message, check-node by check-node, sequentially in iterations, similar to the process of LBP. However, the difference between these algorithms is very small, less than 0.15 dB. Moreover, the difference is almost indistinguishable for irregular codes, as shown in Fig.\ref{fig_7}, where the irregular codes perform 0.4 dB better than the regular ones. This means that the irregular node degree distributions can promote message propagation. Meanwhile, the irregular LDPC codes have better performances than the regular codes, which can deeply improve the message exchanging performance. Thus, in the iterations, the above algorithms can all exchange the extrinsic messages sufficiently to achieve similar performances.

\begin{figure}[!t]
	\centering
	\subfloat[]{\includegraphics[width=1.7in]{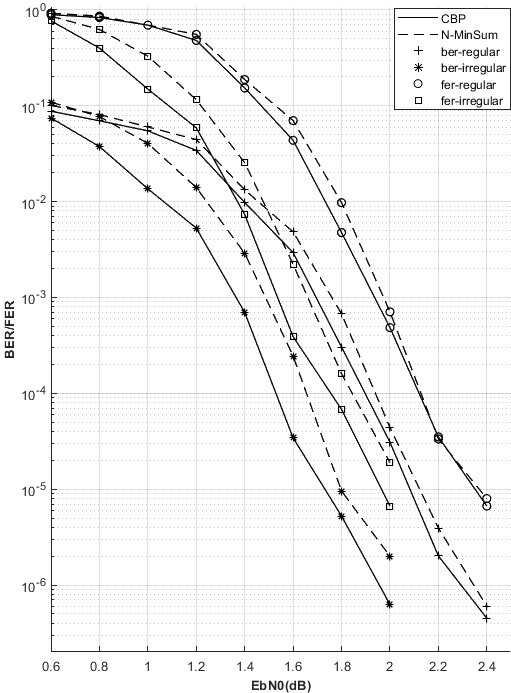}%
		\label{fig_10}}
	\hfil
	\subfloat[]{\includegraphics[width=1.7in]{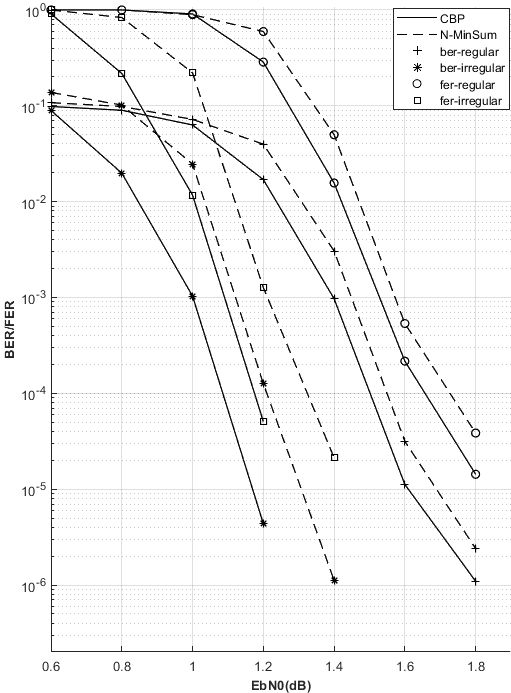}%
		\label{fig_11}}
	\caption{Performance comparisons of the CBP and its normalized min-sum approach. (a)  Length-2048 LDPC codes. (b) Length-8192 LDPC codes.}
	\label{fig_10_11}
\end{figure}

\begin{figure}[!t]
	\centering
	\includegraphics[width=3in,height=2.5in]{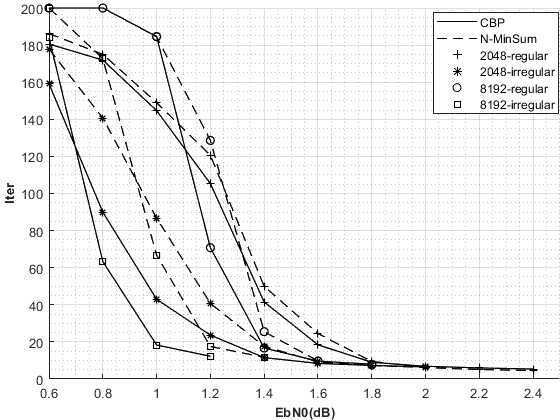}
	\caption{Iteration performance comparisons of the CBP and its normalized min-sum approach for length-2048 and length-8192 LDPC codes.}
	\label{fig_12}
\end{figure}

%\iffalse
Meanwhile, the performance and convergence comparisons for CBP decoding and its normalized min-sum approach are shown in Fig. \ref{fig_10_11} and Fig. \ref{fig_12}. As shown in these figures, 
\iffalse
the result gain loss using the normalized min-sum approach is less than 0.2 dB compared to that of the above \textit{log-tanh} function CBP, 
\fi
the performance loss between the normalized min-sum approach and the above \textit{log-tanh}-based CBP is less than 0.2dB, 
and the number of iterations in the normalized min-sum approach is much larger than that of the CBP in the waterfall area and are almost the same in the low-error-rate area. \tnew{This is because  CBP uses \textit{log-tanh} function to accurately calculate the messages during decoding, but the normalized min-sum approach uses a \textit{normalized-min} function to approximate the \textit{log-tanh} function. Thus, a little performance loss is observed when using the normalize min-sum method. However, it can approximate the \textit{log-tanh} function in CBP with a very small deviation. Thus, both approaches have almost the same number of iterations in the low-error rate area. Thereby, the normalized min-sum approach of CBP is very suitable for hardware implementation.}

\iffalse
This is because  CBP uses \textit{log-tanh} function to accurately calculate the messages during decoding, but the normalized min-sum approach uses a \textit{normalized-min} function to approximate the \textit{log-tanh} function. Thus, little performance loss is observed when using the normalize min-sum method.
\fi

\iffalse
\begin{figure}[!t]
	\centering
	\includegraphics[width=3in,height=3in]{fig10_bfer_minsum_2048}
	\caption{\tdel{Performance comparisons of the CBP and its normalized min-sum approach for length-2048 LDPC codes.}}
	\label{fig_10}
\end{figure}

\begin{figure}[!t]
	\centering
	\includegraphics[width=3in,height=3in]{fig11_bfer_minsum_8192}
	\caption{\tdel{Performance comparisons of the CBP and its normalized min-sum approach for length-8192 LDPC codes.}}
	\label{fig_11}
\end{figure}
\fi

\iffalse
\tdel{Furthermore, the numbers of iterations for CBP decoding and its normalized min-sum approach are shown in Fig. \ref{fig_12}. As shown in these figures, the number of iterations in the normalized min-sum approach is much larger than that of the CBP in the waterfall area and are almost the same in the low-error-rate area. This is because that the normalized min-sum approach has little performance loss, which means that it can approximate the \textit{log-tanh} function in CBP with a very small deviation. Thus, both approaches have almost the same number of iterations in the low-error rate area. Thus, the normalized min-sum approach of CBP is very suitable for hardware implementation.}
\fi

%\fi

\subsection{Calculation Complexity}
The decoding complexity of the LDPC codes includes the convergence iterations $I$, the message updates $Q$ in each iteration, and the calculations $W$ in each message update. The total complexity $T$ can be denoted as
\begin{equation}
	\label{equ_s5e1}
	T = I \times Q \times W
\end{equation} 
These factors will be analysed below.

\subsubsection{Convergence Speed}

\begin{figure}[!t]
	\centering
	\subfloat[]{\includegraphics[width=1.7in]{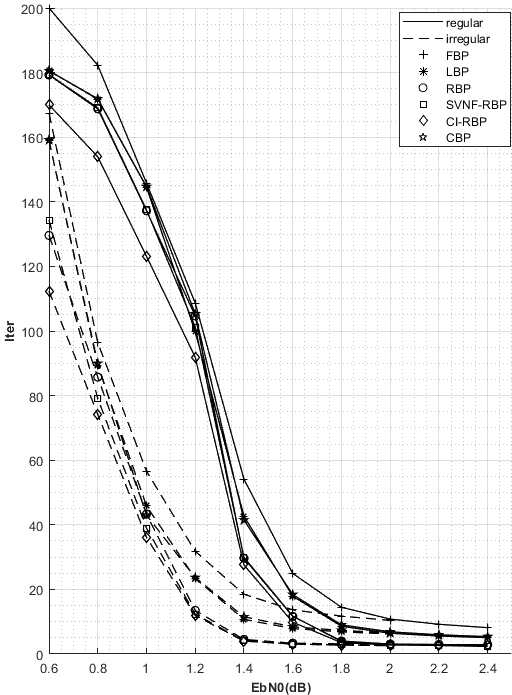}%
		\label{fig_8}}
	\hfil
	\subfloat[]{\includegraphics[width=1.7in]{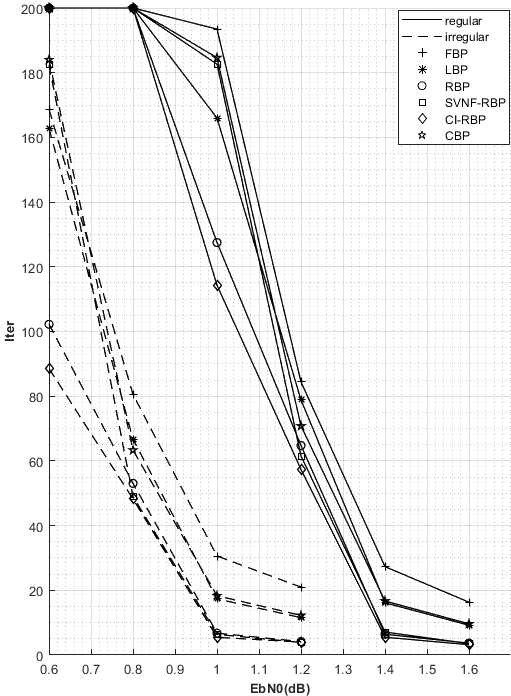}%
		\label{fig_9}}
	\caption{Convergence performance comparisons of the FBP, LBP, RBP, SVNF-RBP, CI-RBP and the proposed CBP scheduling strategies. (a)  Length-2048 LDPC codes. (b) Length-8192 LDPC codes.}
	\label{fig_8_9}
\end{figure}

The convergence speed is measured by the average number of iterations $I$ for  LDPC decoding.  The simulation results for the regular and irregular codes are shown in Fig.\ref{fig_8_9}. From the figures above, we can see that the proposed CBP has twice the convergence speed compared to FBP, similar convergence speed as LBP,  1/2 the convergence speed compared to RBP and SVNF-RBP, and approximately 1/3 the convergence speed compared to CI-RBP. This is because the FBP exchanges messages between phases in a fully parallel manner, which results in one update for each message in each iteration. LBP and CBP exchange messages between layers in a sequential order; thus, messages can be propagated serially in columns. Thus, these approaches have better convergence speed. The RBP and SVNF-RBP approaches can improve the convergence speed through edge-by-edge updating and finish the row and column edge updating in two dimensions in each update. The CI-RBP approach can find the variable-nodes with probable incorrect decisions in RBP scheduling; thus, it has the best convergence speed.

\subsubsection{Updates in Each Iteration}

\begin{table*}[!t]
	\caption{Average Updates in Each Iteration \label{tab:table1}}
	\centering
	\setlength{\tabcolsep}{1.5mm}
	\begin{tabular}{ccccccccc}
		\hline
		Schedules & V2C & C2V & B2V & C2B & Residual & Comparison & Dispatching & CI\\
		\hline
		FBP & $E$ & $E$ & 0 & 0 & 0 & 0 & $E$ & 0\\
		LBP & $E$ & $E$ & 0 & 0 & 0 & 0 & $E$ & 0\\
		RBP & $\sum_i E\lambda_i(d_v^i - 1)$ & $E$ & 0 & 0 & $\sum_{i,j} E\lambda_i \rho_j (d_v^i - 1)(d_c^j - 1)$ & $E(E - 1)$ & 0 & 0\\
		SVNF-RBP & $\sum_i E\lambda_i(d_v^i - 1)$ & $E$ & 0 & 0 & $\sum_{i,j} E\lambda_i \rho_j (d_v^i - 1)(d_c^j - 1)$ & $E(E - 1)$ & 0 & 0\\
		CI-RBP & $\sum_i E\lambda_i(d_v^i - 1)$ & $E$ & 0 & 0 & $\sum_{i,j} E\lambda_i \rho_j (d_v^i - 1)(d_c^j - 1)$ & $E(E - 1)$ & 0 & $E(E - 1)$\\
		CBP & $E$ & 0 & $E$ & $E$ & 0 & 0 & 0 & 0\\
		\hline
	\end{tabular}
\end{table*}

In Table \ref{tab:table1}, the average updates of the different decoding schedules are presented. The number of updates for the FBP, LBP, RBP and SVNF-RBP schedules can be obtained from Table I in \cite{ref20}, while that of the CI-RBP schedule can be obtained from Table I in \cite{ref22}. For each check-belief process in CBP, there are $d_c^j$ B2V updates, $d_c^j$ V2C updates and $d_c^j$ C2B updates. Thus, 	the numbers of B2V, V2C and C2B updates are all $\sum_j (E\rho_j/d_c^j) \cdot d_c^j = E$.
%As shown in Table \ref{tab:table1}, FBP and LBP have the same number of message updates. 
%Meanwhile, RBP, SVNF-RBP and CI-RBP also have the same number of message updates. Additionally, the updates in RBP and SVNF-RBP are scheduled, edge by edge, according to the C2V message-based residual, and the precomputation of the C2V messages is needed. Furthermore, the CI-RBP has a big mount of CI updates to find the variable-nodes with probable incorrect decisions in RBP scheduling \cite{ref22}.
%There are $E$ edges to schedule. In each schedule, $d_v^i-1$ neighbouring V2C messages are updated for a variable-node with degree $d_v^i$, and after the update of each variable-node, $d_c^j-1$ residuals for the neighbouring check-nodes with degree $d_c^j$ are generated. Thus, the number of V2C updates is $d_v^i-1$ times that of the C2V updates, and the number of precomputations is $d_c^j-1$ times that of the V2C updates. As there are $E \lambda_i$ edges with degree $d_v^i$, and $E \rho_j$ edges with degree $d_c^j$, 
%The numbers of V2C updates and precomputations in each iteration for RBP and SVNF-RBP are as shown in Table \ref{tab:table1}. Furthermore, in each schedule, the maximum residual should be found among the $E$ candidates in RBP and SVNF-RBP; thus, there are $E(E-1)$ comparisons. For CBP, in each check-belief processing, 
%there are $d_c^j$ B2V updates, $d_c^j$ V2C updates and $d_c^j$ C2B updates. Thus, 
%the numbers of B2V, V2C and C2B updates are all $\sum_j (E\rho_j/d_c^j) \cdot d_c^j = E$.

\subsubsection{Calculations in Each Update}

\begin{table*}[!t]
	\caption{Average Calculations in Each Update \label{tab:table2}}
	\centering
	\setlength{\tabcolsep}{1.8mm}
	\begin{tabular}{ccccccccc}
		\hline
		Schedules & \tabincell{c}{V2C\\(sums)} & \tabincell{c}{C2V \\(products)} & \tabincell{c}{B2V \\(products)} & \tabincell{c}{C2B \\(products)}  & \tabincell{c}{Residual \\(products)} & \tabincell{c}{Comparison \\(comparision)} & \tabincell{c}{Dispatching \\(selection)} & \tabincell{c}{CI \\(products)} \\
		\hline
		FBP & 2 & 2  & 0 & 0 & 0 & 0 & $max(d_v^i$, $d_c^j)$ & 0 \\
		LBP & 2 & 2 & 0 & 0 & 0 & 0 & $max(d_c^j)$ & 0\\
		RBP & $(d_v^i - 1)$ & $(d_c^j - 1)$ & 0 & 0 & $(d_c^j - 1)$ & 1 & 0 & 0\\
		SVNF-RBP & $(d_v^i - 1)$ & $(d_c^j - 1)$ & 0 & 0 & $(d_c^j - 1)$ & 1 & 0 & 0\\
		CI-RBP & $(d_v^i - 1)$ & $(d_c^j - 1)$ & 0 & 0 & $(d_c^j - 1)$ & 1 & 0 & 2\\
		CBP & 2 & 0 & 1 & 1 & 0 & 0 & 0 & 0 \\
		\hline
	\end{tabular}
\end{table*}

In Table \ref{tab:table2}, the average numbers of calculations for each update of FBP, LBP, RBP, SVNF-RBP, CI-RBP and the proposed CBP strategies are presented. The numbers of sums and products in the V2C updates, C2V updates and precomputation are the node degree minus one in RBP, SVNF-RBP and CI-RBP. However, the numbers of V2C updates and C2V updates are all two because they can be carried out in a row or column scheduling manner. Unfortunately, they need an additional in-row/in-column scheduling process for reserving and dispatching the original messages. Furthermore, in CBP, as described in Algorithm 1, there is one product for the B2V update, two sums (including substrates) for the V2C update and one product for the C2B update.
Specifically, in CI-RBP, there are exponential functions and division operations, which are no less than two products.% not comparable to sum operations.

\subsubsection{Total Complexity}

To determine the total complexity, we set the convergence speeds as 1, 1/2, 1/2, 1/4, 1/4 and 1/6 for FBP, LBP, CBP, RBP, SVNF-RBP and CI-RBP, respectively, as discussed above. By combining the convergence speed and data in Table \ref{tab:table1} and Table \ref{tab:table2}, the total number of calculations in each iteration is determined and shown in Table \ref{tab:table3}. 

\begin{table*}[!t]
	\caption{Total Complexity \label{tab:table3}}
	\centering
	\begin{tabular}{cccccc}
		\toprule[2pt]
		Schedules & Sums & Products & Comparison & Selection\\
		\midrule[1.5pt]
		FBP & $2E$ & $2E$ & 0 & $E \cdot max(d_v^i, d_c^j)$\\ \hline
		LBP & $E$ & $E$ & 0 & $E \cdot max(d_c^j)/2$\\ \hline
		RBP & $\sum_i E\lambda_i(d_v^i - 1)^2/4$ &  \tabincell{c}{$\sum_j E\rho_j (d_c^j - 1)/4 + $ \\ $ \sum_{i,j} E\lambda_i\rho_j (d_v^i - 1)(d_c^j - 1)^2 /4$} & $E(E - 1)/4$ & 0 \\ \hline
		SVNF-RBP & $\sum_i E\lambda_i(d_v^i - 1)^2/4$ & \tabincell{c}{$\sum_j E\rho_j (d_c^j - 1)/4+$ \\ $ \sum_{i,j} E\lambda_i\rho_j (d_v^i - 1)(d_c^j - 1)^2 /4$} & $E(E - 1)/4$ & 0 \\ \hline
		CI-RBP & $\sum_i E\lambda_i(d_v^i - 1)^2/6$ & \tabincell{c}{$\sum_j E\rho_j (d_c^j - 1)/6 + $ \\ $ \sum_{i,j} E\lambda_i\rho_j (d_v^i - 1)(d_c^j - 1)^2 /6+$ \\ $ E(E-1)/3$} & $E(E - 1)/6$ & 0 \\ \hline
		CBP & $E$ & $E$ & 0 & 0\\
		\bottomrule[2pt]
	\end{tabular}
\end{table*}

As shown in Table \ref{tab:table3}, the complexity of FBP, LBP and CBP is $O(E)$, while that of RBP, SVNF-RBP and CI-RBP is $O(E^2/4)$. As $E$ is the number of edges, it is a very big integer. Thus, the complexity of RBP, SVNF-RBP and CI-RBP is much larger than that of 
\iffalse
FBP, LBP and 
\fi
CBP.  Obviously, there are approximately $E$ more sums, $E$ more products and $E\cdot max(d_v^i,d_c^j)$ more selections for FBP than for CBP. There are $E\cdot max(d_c^j)$ more selections for LBP than for CBP. Thus, the complexity of FBP and LBP is larger than that of CBP, too. Altogether CBP has the lowest calculation complexity among the previous introduced strategies.
 
To explicitly illustrate the comparisons, numerical results of Table \ref{tab:table3} are shown in Table \ref{tab:table3x} and Table \ref{tab:table3y} for (3,6) regular LDPC codes and the above ($\lambda(x), \rho(x)$) irregular ones, respectively.
%Here, the product operation can be simplified by the \textit{min}-function, as described in (\ref{equ_s4e18xx}) and (\ref{equ_s4e21xx}). Thus, its complexity is equivalent to two sums. As degrees $d_v^i$ and $d_c^j$ are variables, the selection accesses data in a wide range, and its complexity is consider as 1/4 that of one sum\cite{ref25}. The comparision is indeed a substract, and its complexity is approximate equivalent to a sum, too.
\tnew{As shown in Table \ref{tab:table3x} and Table \ref{tab:table3y}, there are $O(E^2)$ more comparisons and $O(E)$ more products in RBP, SVNF-RBP and CI-RBP than in 
CBP. Meanwhile, the number of sums is much less than that of products and comparisons. As $E$ is usually more than hundreds, the complexity of RBP, SVNF-RBP and CI-RBP would be hundreds of times that of CBP.
On the other hand, there are multiple of $E$ more selections, $E$ more sums and $E$ more products in FBP than in CBP, and multiple of $E$ more selections in LBP than in CBP. Thus, CBP earns a big benefit in terms of calculation complexity.}

\begin{table}[!t]
	\caption{Total Complexity for (3,6) LDPC Codes \label{tab:table3x}}
	\setlength\tabcolsep{4pt} 
	\centering
	\begin{tabular}{cccccccc}
		\hline
		Schedules & Sums & Products & Comparisons & Selections \\
		\hline
		FBP & $2E$ & $2E$ & 0 & $6E$\\
		LBP & $E$ & $E$ & 0 & $3E$ & \\
		RBP & $E$ & $ 13.75E$ & $E(E - 1)/4$ & 0\\
		SVNF-RBP & $E$ & $ 13.75E$ & $E(E - 1)/4$ & 0 \\
		CI-RBP & $0.67E$ & $ E(E/3+8.83)$ & $E(E - 1)/6$ & 0 \\
		CBP & $E$ & $E$ & 0 & 0 \\
		\hline
	\end{tabular}
\end{table}

\begin{table}[!t]
	\caption{Total Complexity for the above ($\lambda(x), \rho(x)$) LDPC Codes \label{tab:table3y}}
	\setlength\tabcolsep{4pt} 
	\centering
	\begin{tabular}{ccccccc}
		\hline
		Schedules & Sums & Products & Comparisons & Selections\\
		\hline
		FBP & $2E$ & $2E$ & 0 & $12E$ \\
		LBP & $E$ & $E$ & 0 & $3E$ & \\
		RBP & $5.04E$ & $ 15.76E$ & $E(E - 1)/4$ & 0 \\
		SVNF-RBP & $5.04E$ & $ 15.76E$ & $E(E - 1)/4$ & 0  \\
		CI-RBP & $3.36E$ & $ E(E/3+10.17)$ & $E(E - 1)/6$ & 0 \\
		CBP & $E$ & $E$ & 0 & 0 \\
		\hline
	\end{tabular}
\end{table}

%As discussed above, CBP consumes only one in hundreds of the calculation resources compared to RBP, SVNF-RBP and CI-RBP, less than 1/3.43=0.29 of the calculation resources compared to FBP, and less than 1/1.71=0.59 of the calculation resources compared to LBP. Thus, CBP is very beneficial in terms of calculation complexity.

\subsection{Memory Consumption}
The memory consumed during decoding is depicted in Table \ref{tab:table4}.  Specially, variable-depth pool uses registers for variable-connections while the others use general memories. \tnew{As shown in Table \ref{tab:table4}, there are $2E - M$ more general cells for RBP, SVNF-RBP and CI-RBP than for CBP. There are $E - M$ more general cells and $P \cdot max(d_v^i, d_c^j)$ registers for FBP than for CBP. In LBP, there are $M$ less general cells, but $P \cdot max( d_c^j)$ more registers than in CBP. It should be pointed out that a register (not including the mux logics) takes up to 10 to 20 times area of a general memory cell \cite{ref28}. Thus, CBP owns the least memory consumption.} 

%\begin{table*}[!t]
%	\caption{Resource Cosumption \label{tab:table4}}
%	\centering
%	\begin{tabular}{c|ccccc|cc}
%		\hline
%		 \multirow{3}{*}{Schedules}& \multicolumn{5}{c|}{block memories} & \multicolumn{2}{c}{distributed registers} \\
%		 \cline{2-8}
%		  & \multirow{2}{*}{LLR} & C2V & V2C & \multirow{2}{*}{residuals} & \multirow{2}{*}{check-beliefs} & calculating & shifting\\
%		  &  & messages & messages &  &  & messages & messages\\
% 		%  & LLR & \tabincell{c}{C2V\\messages} & \tabincell{c}{V2C\\messages} & residuals & check-beliefs & \tabincell{c}{calculating\\messages} & \tabincell{c}{shifting\\messages}\\
%		\hline
%		FBP & $N$ & $E$ & $E$ & 0 & 0 & $P$ & $P \cdot max(d_v^i, d_c^j)$\\
%		LBP & $N$ & $E$ & 0 & 0 & 0 & $P$ & $P \cdot max(d_v^i, d_c^j)$\\
%		RBP & $N$ & $E$ & $E$ & $E$ & 0 & $P$ & $P \cdot max(d_v^i, d_c^j)$\\
%		SVNF-RBP & $N$ & $E$ & $E$ & $E$ & 0 & $P$ & $P \cdot max(d_v^i, d_c^j)$\\
%		CI-RBP & $N$ & $E$ & $E$ & $E$ & 0 & $P$ & $P \cdot max(d_v^i, d_c^j)$\\
%		CBP & $N$ & $E$ & 0 & 0 & $M$ & $P$ & 0\\
%		\hline
%	\end{tabular}
%\end{table*}

\begin{table}[!t]
	\caption{Memory Consumption \label{tab:table4}}
	\setlength\tabcolsep{2pt} 
	\centering
	\begin{tabular}{c|ccccccc}
		\hline
		Schedules & LLR & C2V & V2C & Residual & Check-belief & variable-depth pool\\
		\hline
		FBP & $N$ & $E$ & $E$ & 0 & 0  & $P \cdot max(d_v^i, d_c^j)$\\
		LBP & $N$ & $E$ & 0 & 0 & 0 & $P \cdot max( d_c^j)$\\
		RBP & $N$ & $E$ & $E$ & $E$ & 0 & 0\\
		SVNF-RBP & $N$ & $E$ & $E$ & $E$ & 0 & 0\\
		CI-RBP & $N$ & $E$ & $E$ & $E$ & 0 &0\\
		CBP & $N$ & $E$ & 0 & 0 & $M$ &  0\\
		\hline
	\end{tabular}
\end{table}

%LLRs, $E$ C2V messages, $E$ V2C messages and $E$ residuals.  $P$ calculating pool for temporarily retaining the calculation results, and $P \cdot max(d_v^i, d_c^j)$ scheduling pool for the row/column sheduling, where $P$ is the parallel decoding degree. The number of residuals can be reduced in FBP, saving much memory. \tcolor{In LBP, there are no residuals, and V2C messages are not reserved in memories as they are used as temporarily messges to renew the posterior LLR following (\ref{equ_s2e10}) and (\ref{equ_s2e12}). This further reduces memory consumption.} \tcolor{Moreover, in CBP, shifting messages are eliminated according the check-beliefs propagation process following (\ref{equ_s4e18}) and (\ref{equ_s4e21}), and $M$ additional check-beliefs are needed. Naturally, the number of check-beliefs $M$ is approximately similar to that of shifting messages. However, the block memory for check-beliefs uses less than 1/10 area that the distribued registers for shifting messages, which results in the lowest memory consumption.}

Reducing the variable-depth pool can result in a large improvement. % in terms of implementation. As shown in Table \ref{tab:table4}, the number of registers reduced by shifts is $max(d_v^i, d_c^j)$ times that of message calculation registers. As the degree $d_c^j$ is no less than 6 in applications, more than 6 times the message calculation registers of previous algorithms can be reduced. 
%Indeed, it saves much more registers than 6. 
For example, for the parallel decoding of LDPC codes in 5G NR, assuming a parallel number of $P = 384$, $M = 46P$, the depth of the pool ranges from 3 to 19, and each message is soft quantized by 8 bits \cite{ref23}. During decoding, the CBP can reduce approximately $384 \times 19 \times 8 = 58,368$ registers. Additionally, it would cost $384 \times 46 \times 8 = 141,312$ bits of general memories for reserving check-beliefs, which is equivalent to 14,131 registers. Totally, CBP would reduce about $58,368 - 14,131 = 44,237$ bits of registers, compared to other algorithms. A very large number of memory resources is saved, which will improve the cost significantly.

\section{Conclusion}
In this paper, we have presented an innovative strategy, CBP, based on the check-belief of each check-node. Each check-belief is propagated from one check-node to another check-node through only one variable-node in a recursive manner. This method can strengthen the check-belief sufficiently, ensuring that the corresponding parity check is satisfied and successful decoding is achieved. Compared to previous algorithms employing a large number of cumulative calculations, CBP decoding can renew messages through only two other nodes with no cumulative calculations. This results in reducing calculations and memories for in-row/in-column message scheduling, and earns a low complexity decoding. Simulation results show that the CBP algorithm has no performance loss in contrast with the previous introduced algorithms. The analyses show that CBP consumes only one in hundreds of the calculation resources of RBP, SVNF-RBP and CI-RBP, much less calculation resources than FBP and LBP. Meanwhile, compared with previous algorithms, it can reduce a large number of registers. It has a much lower complexity than the above mentioned algorithms.

{\appendix[Proof of Inclusive Equations]
Here, we note that $\phi(x)$ in (\ref{equ_s2e5}) is a self-reciprocal function, that is, $	\phi^{-1}(x) = \phi(x), \forall x > 0 $.

Define the right part of (\ref{equ_s4e3}) as follows:
\begin{equation}
	\label{equ_ae2}
	Y_n = \phi \left( \sum_{i=0}^{n-1} \phi \left( x_i \right) \right).
\end{equation}

The inverse of (\ref{equ_ae2}) is
\begin{equation}
	\label{equ_ae3}
	\phi^{-1}(Y_n) = \sum_{i=0}^{n-1} \phi \left( x_i \right)
\end{equation}

According to (\ref{equ_ae3}), we can obtain
\begin{equation}
	\label{equ_ae4}
	\begin{split}
		\phi^{-1}(Y_{n+1}) &= \sum_{i=0}^{n} \phi \left(\left| x_i \right| \right) \\
		& = \phi^{-1}(Y_{n}) + \phi(x_n) \\
		& = \phi(Y_{n}) + \phi(x_n)
	\end{split}
\end{equation}

Thus,
\begin{equation}
	\label{equ_ae5}
	Y_{n+1} = \phi \left( \phi(Y_{n}) + \phi(x_n) \right) 
\end{equation}

This proves the recursive function in (\ref{equ_s4e5}).

According to (\ref{equ_ae3}), we can also obtain
\begin{equation}
	\label{equ_ae6}
	\begin{split}
		\phi^{-1}(Y_{n-1}) &= \sum_{i=0}^{n-2} \phi \left(\left| x_i \right| \right) \\
		& = \phi^{-1}(Y_{n}) - \phi(x_{n-1}) \\
		& = \phi(Y_{n}) - \phi(x_{n-1})
	\end{split}
\end{equation}

Thus,
\begin{equation}
	\label{equ_ae7}
	Y_{n-1} = \phi \left( \phi(Y_{n}) - \phi(x_{n-1}) \right) 
\end{equation}

According to (\ref{equ_ae3}), we know that $\phi(Y_{n}) > \phi(x_{n-1})$. However, the result of $\left(\phi(Y_{n}) - \phi(x_n) \right) $ would be a negative value when it is very small. Thus, we usually use (\ref{equ_ae7}) as follows:
\begin{equation}
	\label{equ_ae8}
	Y_{n-1} = \phi \left( \left| \phi(Y_{n}) - \phi(x_{n-1}) \right| \right) 
\end{equation}

This proves the recursive function in (\ref{equ_s4e12}).

\end{document}